\documentclass[sigconf]{acmart}

\begin{document}

\title{MaintainoMATE: A GitHub App for Intelligent Automation of Maintenance Activities}

\author{Anas Nadeem}
\affiliation{
  \institution{\textit{Dept. of Computer Science\\North Dakota State University}}
  \city{Fargo}
  \country{USA}
}
\email{anas.nadeem@ndsu.edu}
\author{Muhammad Usman Sarwar}
\affiliation{
  \institution{\textit{Dept. of Computer Science\\North Dakota State University}}
  \city{Fargo}
  \country{USA}
}
\email{muhammad.sarwar@ndsu.edu}
\author{Muhammad Zubair Malik}
\affiliation{
  \institution{\textit{Dept. of Computer Science\\North Dakota State University}}
  \city{Fargo}
  \country{USA}
}
\email{zubair.malik@ndsu.edu}

\settopmatter{printacmref=false}
\setcopyright{none}
\renewcommand\footnotetextcopyrightpermission[1]{}
\pagestyle{plain}
\keywords{datasets, neural networks, gaze detection, text tagging}

\begin{abstract}
  \textbf{Background: }
  Software development projects rely on issue tracking systems at the core of tracking maintenance tasks such as bug reports, and enhancement requests. Incoming issue-reports on these issue tracking systems must be managed in an effective manner. First, they must be labelled and then assigned to a particular developer with relevant expertise. This handling of issue-reports is critical and requires thorough scanning of the text entered in an issue-report making it a labor-intensive task.
  \textbf{Objective: }
  In this paper, we present a unified framework called MaintainoMATE, which is capable of automatically categorizing the issue-reports in their respective category and further assigning the issue-reports to a developer with relevant expertise.
  \textbf{Method: }
  We use the Bidirectional Encoder Representations from Transformers (BERT), as an underlying model for MaintainoMATE to learn the contextual information for automatic issue-report labeling and assignment tasks. We deploy the framework used in this work as a GitHub application.
  \textbf{Results: }
  We empirically evaluate our approach on GitHub issue-reports to show its capability of assigning labels to the issue-reports. We were able to achieve an F1-score close to 80\%, which is comparable to existing state-of-the-art results. Similarly, our initial evaluations show that we can assign relevant developers to the issue-reports with an F1 score of 54\%, which is a significant improvement over existing approaches.
  \textbf{Conclusion:}
  Our initial findings suggest that MaintainoMATE has the potential of improving software quality and reducing maintenance costs by accurately automating activities involved in the maintenance processes. Our future work would be directed towards improving the issue-assignment module.
  More specifically, we plan to study adding external features i.e. activity-based features, developer profile-based features to evaluate if incorporating these features improve the overall results of the issue assignment module.
\end{abstract}

\maketitle

\section{Introduction}
Software development projects rely on the use of issue tracking systems as a unified source of communication of all software maintenance activities. In large scale projects their adoption no longer remains a choice and becomes a necessity as teams working on these projects may be dispersed and distributed in several geographical regions \cite{murphy2004automatic}. GitHub, a widely used \cite{kalliamvakou2014promises} source code hosting platform associates a general-purpose issue tracking system to every project and is used by industry projects such as .NET \footnote{\url{https://github.com/microsoft/dotnet/issues}}, TensorFlow \footnote{{\url{https://github.com/tensorflow/tensorflow/issues}}}, and Kubernetes\footnote{{\url{https://github.com/kubernetes/kubernetes/issues}}}. Under usual circumstances, any user can report an issue onto these issue tracking systems. Effective software maintenance is a measure of how well these reports are managed i.e., incoming reports must be addressed with a resolution quickly \cite{murphy2004automatic}.
The anatomy of a newly created issue-report on an issue tracking system typically comprises of a title and a description. Optionally, these systems allow assigning a label to these issue-reports which indicates the nature and severity of the issue report (bug, enhancement, question). However the issue-reports are rarely labeled by the submitter. Once this newly created issue report is received, it requires further handling before anyone can start working on its resolution: (1) An incoming issue-report must be classified into one of the available labels to allow prioritization of these issues (2) An incoming issue report must be assigned to a developer having relevant expertise to provide a quick resolution of the issue-report \cite{anvik2006should}.

The common practice followed by the software development community is to manually assign labels and a developer to these issue reports. These assignments usually rely on identifying relevant features through textual data of the issue report. A previous study by Lin et al. \cite{bug_assignment_chinese} found textual data to be most useful during the triaging process as compared to non-textual information. Triaging process refers to actions undertaken in order to prioritize and assign the reports to developers. The team leader identifies the context of the issue-report through the text and assigns the issue-report to the relevant developer. Although this type of handling is not a direct part of the actual issue-report resolution, it has a significant impact on the software maintenance process and the resolution time of the reports. While the manual process of assignment seems feasible for small scale projects, it is insufficient for the maintenance of large scale projects for several reasons: (1) Assigning issue-reports to developers becomes a non-trivial and time-consuming task \cite{anvik2006should} since the frequency of issue reports is very high and the reviewer of the report is required to thoroughly read through the text (2) With changes in a software team, it becomes difficult to keep track of expertise of every team member while assigning these issues \cite{matter2009assigning}. Mishandling of these issue-reports might result in poor execution of software maintenance activities. For instance, a bug that was incorrectly labeled as a feature might slip into production. Similarly, a developer who is experienced at working on the back-end might not be the best fit to work on issues related to the User Interface.

Existing approaches to solving these problems primarily rely on keyword-based approaches \cite{ahmed2014predicting, alenezi2013efficient, antonio, anvik2006should, dedik2016automated, FAZAYELI2019585, KALLIS2021102598, kevic2013collaborative, matter2009assigning, murphy2004automatic} and suffer from a high false positive and a high false negative rate. To address the aforementioned issues, we use a transformer-based neural network called RoBERTa \cite{roberta}. RoBERTa has the ability to understand the context, we fine-tune this language representation model to use as an underlying model for our framework MaintainoMATE. MaintainoMATE has the following two modules: (1) Issue-Report Labelling: which categorizes the issue-report into its respective category i.e. `Bug-Report', `Enhancement', `Question' (2) Issue-Report Assignment: which assigns the issue-report to the relevant developer. To evaluate Issue-Report Labelling module, we randomly sampled 55,000 issue-reports associated with the top 200 repositories across 55 popular languages. We were able to achieve promising results with an F1-Score nearing 80\% which is comparable to the existing state of the art results. In a similar manner, to evaluate our Issue-Report Assignment module, we trained the model on issue-reports along with the assigned developers of `TensorFlow' repository on GitHub. Our model shows promising results i.e. 54\% which is a significant improvement over the existing state of the art results. 

Previous research works \cite{ahmed2014predicting, alenezi2013efficient, antonio, anvik2006should, dedik2016automated, FAZAYELI2019585, KALLIS2021102598, kevic2013collaborative, matter2009assigning, murphy2004automatic} aimed to automate various tasks of software maintenance, lack in industry adoption which is a result of scarcity of practical tools available to the software industry. Our research is aimed towards bridging this gap between industry expectations and scientific methods. To the best of our knowledge, our tool undertakes a novel approach towards being a full suite of modules curated for intelligently automating various steps involved in software maintenance. We release MaintainoMATE \footnote{\url{https://github.com/apps/maintainomate} - Made private for blind review as GitHub apps show name of the developer} as a GitHub application which allows its easy integration with any software repository on GitHub.

In this paper we claim the following contributions:
\begin{itemize}
    \item We present an automated software maintenance framework `MaintainoMATE'. It comprises of two components: `Issue-Report Labelling', and `Issue-Report Assignment'.
    \item We present our complete framework for Issue-Report Labelling component, which performs the classification of issue-reports into their respective categories i.e. `Bug-report', `Question', and `Enhancement'. Our approach has comparable results with the state-of-the-art studies with an F1-score of 80\%.
    \item We discuss the initial findings of utilizing a novel approach for Issue-Report Assignment component, which assigns issue-reports to the respective developers based on their experiences. Our model shows promising results i.e. 54\% F1 score, which is a significant improvement over the existing state of the art results.
    %\item We set the future direction of our research, which is focused on adding additional features i,e, developer profiling to our tool to further improve our results for effective software maintenance.
\end{itemize}

\textbf{Organization.} The paper is structured as follows: Section \ref{sec:related_Work} discusses the relevant previous research works. Section \ref{sec:maintainomate} discusses the different components of the our proposed MaintainoMATE framework. Section \ref{sec:results} discusses the evaluation of our approach. Section \ref{sec:threats} discusses threats to the validity of our work. Section \ref{sec:future_work} discusses the future avenues of our work. Finally, section \ref{sec:conclusion} concludes our paper. 

\section{Related Work}
\label{sec:related_Work}
This section discusses the previous studies focused on solving such problems. Primarily, this section presents the previous studies on (1) Issue-Report Labeling (2) Issue-Report Assignment (3) Transformer (4) Industry tools in software maintenance. 
%`Issue-Report Labelling', and `Issue-Report Assignment'.
\subsection{Related Work: Issue-Report Labeling}

Several research works have presented methods ranging from using keyword-based approaches to machine learning-based models. We discuss such research works along with their limitations in order to lay the grounds for our research.

%Antoniol et al.\cite{antonio} proposed a key-word based approach in order to distinguish bug-issues from non-bug issues. They manually labeled 1,800 issues associated with repositories of Mozilla, Eclipse, and JBoss and further fed these issues to Alternating Decision Trees, Naive Bayes, and Logistic Regression-based models resulting in up to 82\% correct results. Our work proposes a multi-label method to classify these issues into three mutually exclusive labels. Additionally, we evaluate our results on a larger set of issues from a diverse set of repositories. 

Kallis et al. \cite{KALLIS2021102598} presented a tool i.e. TicketTagger to classify the GitHub issue-reports. They utilized a fastText \cite{joulin2016bag} based model to classify the GitHub issue into their respective categories i.e. bug, enhancement, and question. Over these three categories, they achieved a F1-score of 82\%. However, they have used a multi-class setting to solve the problem. In multi-class setting, one issue-report can belong to a single label at a time. Our work used similar labels as used by Kallis et al. \cite{KALLIS2021102598}. However, we solve this problem as multi-label classification problem, where an issue-report can be assigned more than one label at the same time.

Fan et al. \cite{fan} proposed a machine learning based approach for classifying the issue-reports. They evaluated their approach on over 252,000 issue reports from 80 popular GitHub projects. Further, they evaluated four traditional machine learning methods i.e. Naive Bayes, Logistic Regression, Random Forest Tree, and Support Vector Machine. Our work, in contrast, leverages the off-the-shelf transformer-based model which achieves state-of-the-art results without requiring a large number of training set.

%Chawla et al. \cite{chawla} proposed an automated approach to categorize the type of a issue-reports. They utilized a Fuzzy Logic-based system to classify the issue types and evaluated their results on issue-reports from HTTPClient, Jackrabbit, and Lucene. They were able to achieve a F-1 score of 0.83, 0.79, 0.84 for the three repositories, respectively. However, they evaluated their model on results on a limited number of well-maintained repositories which might be misleading.

Fazayeli et al. \cite{FAZAYELI2019585} used the traditional machine learning-based classification methods to categorize the GitHub issue-reports. However, they only evaluated their approach using one repository i.e. 'git-for-windows'. Moreover, they solved the problem as a binary classification problem with bug and non-bug labels. Our framework assigns multiple labels to a GitHub issue-report at a time. Also, we evaluated our approach on a dataset consisting of multiple projects from various programming languages.

Previously discussed studies suffer from a high rate of false-positives as they are utilizing keyword-based features to categorize the issue-reports, The primary reason for such a high error rate is lack of consideration of contextual information of the text during classification. Also, these studies solved the issue classification problem in a multi-class classification setting. In real-world scenarios, an issue can be associated with more than one label at a time and should be solved in a multi-label classification setting.

%Issue Assignment
\subsection{Related Work: Issue-Report Assignment}

Assigning issues to developers based on their expertise, commonly referred to as bug triaging. Bug triaging has received interest from numerous researchers in the past. Previous researches in this area have mainly relied on key word-based approaches. The majority of these research works solely rely on textual data of the issue report \cite{anvik2006should, dedik2016automated, alenezi2013efficient, murphy2004automatic} while some of them attempt to complement the data from other sources such as developer source code commits \cite{matter2009assigning}, and contributions on Q\&A platforms i.e. StackOverflow \cite{crowdsourcedtriage}. 

%Efforts were made to study automatic bug triaging using text categorization by Curbranic, D., and Murphy, G. \cite{murphy2004automatic}. They used a Naive Bayes classifier and achieved 30\% accuracy on issue reports from the Eclipse project. 

%Anvik et al. presented a semi-automated approach to recommend a small number of developers to assign to the incoming issue reports\cite{anvik2006should}. They used textual data of the issue report with three classifiers namely Support Vector Machines (SVM), Naive Bayes and C4.5. They found SVM to be performing the best and achieved precision levels of 57\% and 64\% on the Eclipse and Firefox project. Our bug assignment subsystem is fully automated and shows promise for better generality.

Kevic et al. \cite{kevic2013collaborative} utilized a multi-phased approach for their analysis. In addition to the textual similarity of issue reports, they analyzed the changeset associated with the issue report. For any incoming issue-report, they first compute the tf-idf feature vectors and then compute the closeness between those vectors based on cosine similarity to recommend the developers who fixed similar reports. They presented a prototype tool that for a given report recommends a list of developers and is aimed to be used in scrum-based meetings. In comparison, our approach is targeted towards a fully automated solution for the assignment task.

%Lin et al. \cite{bug_assignment_chinese} studied the feasibility of bug assignment automation using SVM on Chinese Bug Data. They that existing approaches give similar results even when mapped to reports containing Chinese text. Their key contributions include utilizing both textual data as well as non-textual information of the report for the classification task. Their results concluded the textual data contained in an issue report to be the most efficient in assigning the bug reports to respective developers.

Matter et al. \cite{matter2009assigning} presented a keyword-based recommender system that models the expertise of the developers using the vocabulary used in their source code commits. Further, they compared the commits' vocabulary to the vocabulary used in the issue-reports in order to assign them to developers. Their evaluations on the Eclipse project achieved 33.6\% precision. In comparison, we rely on textual data included in past issue reports only.

Alenezi et al. \cite{alenezi2013efficient} explored the use of term selection methods in order to identify discriminating terms used in a report to assign the issues. Their results found X2 term selection method to outperform their baselines by over 20\% on some projects.

%Transfomer
\subsection{Related Work: Transformers}

%Attention-based networks derived from human intuition have resulted in significant improvement in various natural language processing tasks\cite{attention_nlp}. Attention-based networks are capable of focusing on context-relevant details in a given text while ignoring irrelevant keywords. Bahdanau et al.\cite{bahdanau2016neural} were the first to introduce an attention mechanism for the purpose of machine translation, commonly referred to as additive attention. 

Attention-based networks derived from human intuition have resulted in significant improvement in various natural language processing tasks\cite{attention_nlp}. This led to wide adoption of self-attention based Transformers \cite{vaswani2017attention} models. Pre-trained transformers are trained on a large natural language corpus and further, they can be fine-tuned for the downstream NLP tasks such as machine translation, and text classification.

In recent work, Devlin et al.\cite{bert} introduced Bidirectional Transformers for Language Understanding (BERT) which leverages the Transformer-based architecture. BERT is pre-trained on 800M words of BooksCorpus and 2500 million words of English Wikipedia. Such pre-training enables the BERT to be familiar with vast the range of words in the English language vocabulary. There are several BERT variants that differ in architecture such as the number of input layers, hidden layers, self-attention heads, and parameters used. Liu et al.\cite{roberta} proposed a BERT variant `RoBERTa' which is a robust and optimized variant of BERT. They proposed several design changes such as removing the next sequence prediction objective and changing the masking pattern based on the data itself. They show that with all such changes, and pre-training the model for a longer duration and in large batches result in substantial improvement in results on benchmark dataset such as GLUE, SQuAD and RACE.

%Tooling
\subsection{Related Work: Industry Application of Existing Research}
Despite the abundance of research in mining and analysis of data available on issue tracking systems, there is a lack of industry adoption of these researches. We believe that this scarcity in the practical adoption of these methods is due to two reasons 1) The approaches proposed in the previous works are not generalizable for industry use 2) There is a lack of availability of industry-level tools in order to evaluate these methods for large-scale adaptability.

We found only a few practical adoptions of the scientific methods. TicketTagger \cite{KALLIS2021102598} is a tool proposed by Kallis et al. which labels issues reported on software repositories. Similarly, Develect is a prototype tool \cite{matter2009assigning} which automates the assignment of these issues. Katja et al. \cite{kevic2013collaborative} proposed a semi-automated prototype to be used in collaborative meetings to assign these reports to the developers. Nevertheless, the discussed tools are either semi-automated i.e., require human interaction, and lack real-world adoption due to lack of generalizability.
%\section{Methodology}
\section{MaintainoMATE: Approach and Assembly}
\label{sec:maintainomate}
This section presents individual components of our tool called MaintainoMATE, a comprehensive system aimed at intelligently automate various steps involved in software maintenance. Our system consists of the following components: (1) Issue-Report Assignment Labelling, which classifies the issue-reports into their respective categories i.e. Bug, Enhancement, Question etc. (2) Issue-Report Assignment, which automatically assigns the issue-reports to the relevant developer. We briefly highlight the different components of MaintainoMATE in this section and explain details of each component in the subsequent subsections.

%Subsequent subsections highlight our approach towards individual components of MaintainoMATE.

\subsection{Issue-Report Labelling}
\label{issue_classification}
This section presents how our Issue-Report Labeling module automates the labeling of issue reports into respective categories to enable an effective software maintenance process.

\textbf{\textit{Data Collection and Transformation:}}
We used GitHub as our data source. We identified top 55 programming languages using IEEE Spectrum programming languages ranking list \cite{languageranking}. We further fetched 200 popular repositories of each language. We identified the projects by their primary language. Here, we considered `star' count of the repository as the popularity measure \cite{repopopularity}. Finally, we fetched the issues-reports of each GitHub repository along with the title, body, labels etc. 

We extracted the title and body of the issue and concatenated them together. We filtered our GitHub issue-reports into three categories i.e. bug-report (bug or fault report), enhancement (request for enhancement of the project), and question (question regarding the project). These labels are consistent with the prior work by Kallis et al.\cite{KALLIS2021102598}. Here each label is non-exclusive in nature. For instance, an issue can be labeled as bug-report and enhancement at the same time. In total, we were able to fetch 1,166,107 issues. The substantial amount of the unlabelled issue-report i.e. 509,090, provides us the opportunity to provide an effective solution to categorize the issue-reports. We used GitHub Search and Data API \footnote{\url{https://docs.github.com/en/rest}} to fetch this dataset. This dataset is available for public use \footnote{\url{https://zenodo.org/record/5110986}}.

\textbf{\textit{Model:}} Pre-trained tansfomer based model has the ability to extract the detailed contextual relationship of the words from the text, as they are trained on large natural language corpus. We used one such off-the-shelf model `RoBERTa' \cite{roberta}, which is an optimized version of BERT \cite{devlin2018bert}. The reason behind choosing RoBERTa over BERT is its improvement in results on various benchmarks datasets i.e. GLUE, RACE, and SQuAD as compared to BERT. We choose to use `roberta-base' variant of RoBERTa which consists of 12 transform layers, 768 hidden layers, and 12 self-attention heads. We treated our issue-report labelling problem as multi-label classification problem, where an issue-report can have more than one label at a time. We used simpletransformers \footnote{https://github.com/ThilinaRajapakse/simpletransformers/} to create our model.

\textbf{\textit{Model Training:}} To train our model we randomly sampled 55,000 (approx) issue-reports. Further, we divided the dataset into 80\% and 20\% training and testing dataset split respectively. We trained our model for 5 epochs with a learning rate of 4e-05, a maximum sequence length of 128, and a batch size of 8. We train our model on Google Colab \footnote{\url{https://colab.research.google.com/}}, a free Jupyter notebook based environment that allows to run scripts. It took an hour to train the model with 12 GB RAM, Tesla T4 14 GB GPU and 2 Intel Xeon CPUs @ 2.2GHz specification.

%\begin{figure*}[htbp]
%\centerline{\includegraphics[width=\textwidth]{figures/issues_combined.png}}
%\caption{A demonstration of how BERT focuses important details in issue text to identify issue reports with similar modules and assign them to same developer}
%\label{fig:bert_assign_demo}
%\end{figure*}

%Ans work on the following section "Issue Assignment"

\subsection{Issue-Report Assignment}

This section presents the methodology used in Issue-Report Assignment module of MaintainoMATE. Issue-report assignment module is to assign the issue-reports to the respective developers based on the previously assigned issues.

%The Issue Assignment module of MaintainoMATE also utilizes a similar pre-trained model and is aimed at assigning issues to respective developers with knowledge and expertise required to provide a quick resolution. In this section we highlight how we MaintainoMATE intelligently automates the issue assignment task.

\textbf{\textit{Data Collection and Transformation:}} For issue-report assignment module we sampled 4,500 issue reports (approx) along with the assignee developers from TensorFlow project on GitHub, which is a top ranked \cite{repopopularity} industrial project. To account for software team changes, we carefully sample reports from developers that have had at least 50 issue reports assigned in the year 2021. This reduces the probability of sampling any developers are not working on the project any longer. Our analysis show that 14 developers actively maintain the TensorFlow project. This dataset\footnote{\url{https://zenodo.org/record/5110986}} is also made available to the public as a means of enabling future analysis 

\textbf{\textit{Model:}} We treated the issue-report assignment as a multi-class classification problem where an issue-report can only belong to a single label at a time. Here, the issue-reports textual features i.e. title, description are fed as features while the developers assigned to these issue-reports are the labels of the model. We used the `roberta-base' model similar to the one used in section \ref{issue_classification}.% Figure \ref{fig:bert_assign_demo} demonstrates how our attention-based model looks at issue reports to focus on important details.

%We map the issue assignment task as a multi-class classification problem where the developers who work on the project are the labels. The framework underneath utilizes a similar model i.e., RoBERTa \cite{roberta} as used in the issue classification task. Software projects commonly have designated developers working on specific modules. Using an attention-based framework enables our model to identify which areas a developer typically works on resulting in efficient resource allocation. Figure \ref{fig:bert_assign_demo} demonstrates how our attention-based model looks at issue reports to focus on important details.%We use `roberta-base' which consists of 12 transform layers, 768 hidden layers, and 12 self-attention heads. However, opposed to the issue classification problem we solved the issue assignment problem as multi-class classification problem. 

\textbf{\textit{Model Training:}} 
To train our model, we divide our dataset into 80\% training and 20\% testing datasets. Further, we feed these issue-reports into multi-class RoBERTa classifier and train it for 5 epochs with similar hyper-parameters as discussed in section \ref{issue_classification}. We also used Google Colab \footnote{\url{https://colab.research.google.com/}} to train our model and it took us an hour to train the model.

\subsection{Deployment}

\begin{figure}[htbp]
\centerline{\includegraphics[width=0.5\textwidth]{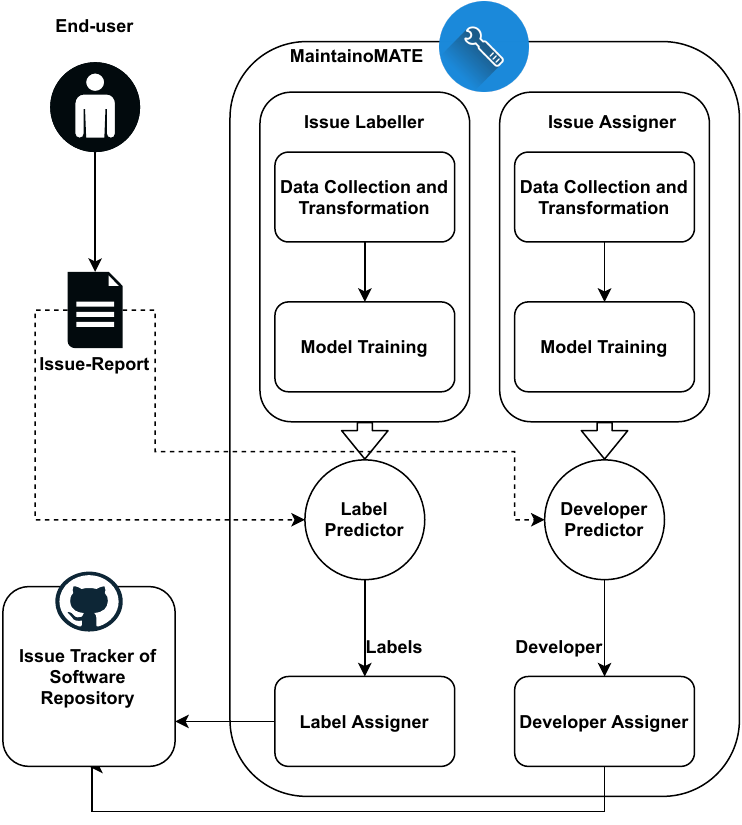}}
\caption{Upon submission of a new report by a user, MaintainoMATE automatically uses the trained models to predict labels and an expert developer to resolve the issue, and finally assigns it back to the report on the issue tracker}
\label{fig:overview}
\end{figure}

MaintainoMATE is deployed as a GitHub application that allows easy integration of the tool in any software project hosted on GitHub. Figure \ref{fig:overview} demonstrates an overview of our approach and the deployment. This GitHub application consists of the following two components: (1) Issue-Report Labelling Component (2) Issue-Report Assignment Component. 

\begin{figure}[htbp]
\centerline{\includegraphics[width=0.5\textwidth]{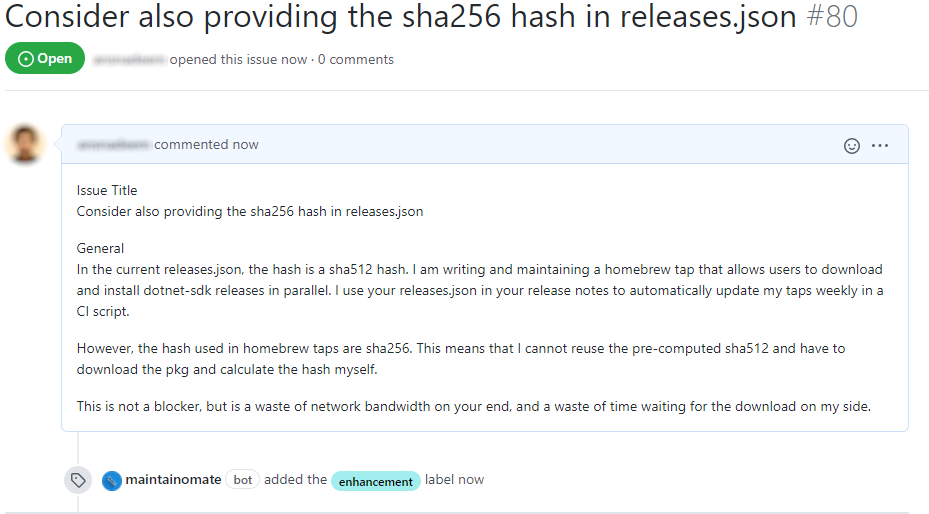}}
\caption{A demonstration of MaintainoMATE, automatically assigning label to a report replicated from a real issue}
\label{fig:autolabel}
\end{figure}

\textbf{\textit{Issue-Report Labelling Component}}
The Issue-Report Labelling Component of MaintainoMATE automatically assigns labels to issue reports on a repository having the tool installed.
When integrated, any new issue is automatically assigned a label out of Bug, Enhancement or Question. Figure \ref{fig:autolabel} shows a demonstration of our app assigning a label to a newly created report which was replicated from a real issue report\footnote{\url{https://github.com/dotnet/core/issues/3407}}.

%\textbf{\textit{Automatic Issue Classifier}}
%Issue classification is deployed as a GitHub application called Automatic Issue Classifier (AIC). AIC application can be integrated with any GitHub project and has the ability to automatically classify the issue-report into its respective label. AIC is a Flask \footnote{\url{https://flask.palletsprojects.com/en/2.0.x/}} based application and is deployed on a WSGI server \footnote{\url{https://wsgi.readthedocs.io/en/latest/index.html}}. AIC application can be integrated with GitHub repository by accessing it through the application page \footnote{\url{https://github.com/apps/automatic-issue-classifier}}. After landing on the application page, click on the `Install' button and select the GitHub repository you want to associate it with. Figure \ref{fig:aic_flow} explains the working flow of the application.

\textbf{\textit{Issue-Report Assignment Component}}
The Issue-Report Assignment Component, when completed would be able to automatically assign experienced developers to issue reports. Although the work on this component is under progress, we have empirically studied and presented a baseline approach with our initial results.
%Automatic Issue Assignment (AIA) is the deployed GitHub application of the issue assignment component of MaintainoMATE system. The work on AIA is still under progress. AIA would have the ability to assign the newly created issue-report to the appropriate developer based on the nature of the issue-report.

%\subsection{Dashboard}
\section{Results and Discussion}
\label{sec:results}
This section discusses the evaluation of results of our framework. We used the following metrics to evaluated our framework:

\begin{itemize}
    \item \textbf{Precision} is the ratio of true positives to the sum of true positives and false positives. A precision value closer to 1 is the most desirable.
    
    \item \textbf{Recall} is the ratio of true positives to the sum of true positives and false positives. A recall value of 1 indicates the best performance. 
    
    \item \textbf{F1-Score} is defined as harmonic mean of precision and recall. F1-Score closer to 1 is the most preferable.
\end{itemize}

\textbf{\textit{Issue-Report Labeling: }}The evaluation of the issue labeling module over a large data set of issue-reports from GitHub gives us an F1-Score of 81\%, 74\% and 80\% for bug, enhancement, and question labels respectively which is comparable to previous studies. For instance, Kallis et al.\cite{KALLIS2021102598} reported F1-Scores of 83.1\%, 82.3\%, 82.5\% for bug, enhancement and questions category respectively. The slight degradation is expected as we proposed a multi-label classification approach as compared to the multi-class approach presented by Kallis et al.\cite{KALLIS2021102598}.  Table \ref{tab:issue_classification_results} gives the complete evaluations of our results.

\textbf{\textit{Issue-Report Assignment: }} The evaluation of our initial approach to the issue assignment task over the issue-reports from the TensorFlow project shows promising results. Our studies show significant improvement to existing keyword-based approaches. For instance, the approach used by Anvik et al.\cite{anvik2006should} assigns developers to issues with precision score of 64\% however they suffer from low recall score of 10\%. Our approach results in comparable precision values while resulting in a significantly improved recall of 52\%, precision of 59\%, and F1-Score of 54\%. Table \ref{tab:assign_results} shows the evaluations of our results.

%\subsection{Issue Classification}

\begin{table}[]
\centering
\begin{tabular}{l|llll}
\hline
 & Precision & Recall & F1-Score \\ \hline
Bug & 81\% & 81\% & 81\% \\
Enhancement & 78\% & 72\% & 74\% \\
Question & 79\% & 81\% & 80\% \\
Macro-Average & 79\% & 78\% & 78\% \\ \hline
\end{tabular}
\caption{Evaluation results of our label assignment study}
\label{tab:issue_classification_results}
\end{table}

%\subsection{Issue Assignment}

\begin{table}[]
\centering
\begin{tabular}{ll}
\hline
  Metric & Score \\ \hline
  Precision & 59\% \\
  Recall & 52\% \\ 
  F1 Score & 54\% \\ \hline
\end{tabular}
\caption{Evaluation results of our developer assignment study}
\label{tab:assign_results}
\end{table}
\section{Threats to Validity}
\label{sec:threats}
This section highlights some threats in validating our work.
\begin{itemize}
    \item \textbf{Bias:}
    We evaluate MaintainoMATE over dataset containing popular repositories of well-maintained projects. 
    This might induce a popularity bias in our results since some of the non-popular repositories might not be well maintained.
    \item \textbf{Cold Start Problem:} The current methodology for issue assignment component of MaintainoMATE requires a repository to have a number of issue-reports in order to learn what patterns are followed in assigning issues-reports to the developers.
    \item \textbf{Language Imbalance:} Our dataset implies some languages are more popular on GitHub than others resulting in our dataset being dominated by issue reports from issue trackers of repositories of these popular languages.
\end{itemize}

\section{Future Work}
\label{sec:future_work}
The future work of our research would be directed towards improving MaintainoMATE and adding more use-cases for our GitHub application. Specifically, we would work on: (1) Issue-Report Labelling (2) Issue-Report Assignment (3) Incorporating Other Use-cases

\subsection{Future Work: Issue-Report Labelling}

\textbf{\textit{Adding More Labels:}} Although our labelling module is capable of categorizing bugs, questions and enhancement with high accuracy. We further plan to explore other default GitHub labels such as wont-fix, help-wanted, duplicate and analyze if they can be incorporated into MaintainoMATE.

\subsection{Future Work: Issue-Report Assignment}
%evaluate the best configuration of each algorithm on a larger dataset of discussions from SO and other textual artefacts (e.g., issues and code reviews), and also consider other topics that are discussed on these forum\

\textbf{\textit{Hyperparameter Tuning:}} We plan to further tune the hyperparameters to figure out the best configuration of our transformer based model, which would help us to improve the results

\textbf{\textit{Developer Profiling}}
Currently, we used issue-report textual features to assign the developers to their respective issue-report. In future, we plan to add other features such as activity-based features developers profile based features etc. Following the feature set we are planning to incorporate into our model:

\begin{itemize}
    \item \textit{Activity-based Features:} This set of features would include features related to the developers' previous interactions with source-code files and with resolving issue-reports. 
    \item \textit{Profile-based Features:} This category of features would contain all the features related to developer's programming activities and their correlation with the issue-reports. For instance, we can include the developers' commit messages. Commit messages can be a rich source for profiling developer expertise as the language used in commit messages is closer to that used in issue-reports \cite{alkhazi2020learning}.

\end{itemize}

%Activity-based features: commit messages->issue scenes

%Location-based features.
%Profile-based features

%The issue assignment module relies on a transformer-based model for assigning issue reports to respective developers, we plan to explore a more advanced \textbf{Metric Learning based Methodology} for this task. The aim of `Metric Learning' is to learn a representation function that maps instances into embedding vectors. Such that similar embedding vectors are close to each other and dissimilar embedding vectors are farther away in the embedding space. We plan to use deep metric learning using character-level convolutional neural network to learn discriminating features from the issue-report assigned to each of the developer. Specifically, we would be training our network using triplet loss function, contrastive loss function, and lifted-structured loss. The we would extract embedding vectors from the trained CNN and further feed them to the supervised classification model such as nearest neighbour classifier. The reason behind utilizing the nearest neighbour classifier is it lazy earning ability i.e. we do not need to retrain it on adding new samples.

\subsection{Future Work: Incorporating Other Use-cases}

%\begin{itemize}
    %\item \textbf{Report and Anlytics:}
In addition to automating the labelling and assignment of the issue-reports, we plan to explore more applications of automation in software maintenance processes. We plan to incorporate a real-time dashboard which would provide real-time analytics to the project mangaer such as number of bug-report issues, number of issue-reports assigned to a particular developer etc. We plan to generate various reports aimed at providing maintenance analytics for various team roles.
    
    \begin{itemize}
    \item\textit{Developer Report:} The developer report can show the number of issue-reports along with the categories assignment to the developer. The primary goal is to analyze the maintenance activities across the development team, e.g., who has been assigned to fix the bug, and who has been assigned to implement a new feature.
    \item\textit{Project Report:} The project report can show the categorized issue-reports along with source code files affected by them across the project. The goal is to provide holistic real-time software maintenance surveillance of the project to managers. 
    \end{itemize}
    
%\end{itemize}

\section{Conclusion}
\label{sec:conclusion}
We have proposed a MaintainoMATE framework that aims to automate various tasks for maintenance of a software project. We have released this framework as GitHub application and can be easily integrated with any GitHub repository. MaintainoMATE is capable of: (1) categorizing an issue-report into its respective category (2) assigning an issue-report to the relevant developer. MaintainoMATE is fully-capable of automatically assigning labels to the issue-reports to identify the nature of these reports as bugs, enhancements or question with state-of-the-art results nearing an F1 score of 80\%. We have also presented initial results from our transformer-based model for the automatic bug assignment task i.e. 54\% F1 score and have discussed our vision towards improving these results. The goal of this work is to promote real-life adaptation of software maintenance model for software development projects. MaintainoMATE would help to improve overall software quality by allowing efficient resource allocation while also lowering maintenance costs by automating various tasks where manual effort is required. Future of research would be directed towards the improvement of the current modules as well as incorporating other automation techniques to facilitate software maintenance by adding more functionality such as report generation and real-time analytics of the project using a dashboard. 

%Automation using the machine learning techniques has a wide-range of practical applications in the area of software maintenance and the future of MaintainoMATE lies on exploring and incorporating these applications.

\bibliographystyle{ACM-Reference-Format}
\bibliography{main}


\begin{thebibliography}{23}

%%% the Web address.

\ifx \showCODEN    \undefined \def \showCODEN     #1{\unskip}     \fi
\ifx \showDOI      \undefined \def \showDOI       #1{#1}\fi
\ifx \showISBNx    \undefined \def \showISBNx     #1{\unskip}     \fi
\ifx \showISBNxiii \undefined \def \showISBNxiii  #1{\unskip}     \fi
\ifx \showISSN     \undefined \def \showISSN      #1{\unskip}     \fi
\ifx \showLCCN     \undefined \def \showLCCN      #1{\unskip}     \fi
\ifx \shownote     \undefined \def \shownote      #1{#1}          \fi
\ifx \showarticletitle \undefined \def \showarticletitle #1{#1}   \fi
\ifx \showURL      \undefined \def \showURL       {\relax}        \fi
% invisible to TeX
\providecommand\bibfield[2]{#2}
\providecommand\bibinfo[2]{#2}
\providecommand\natexlab[1]{#1}
\providecommand\showeprint[2][]{arXiv:#2}

\bibitem[\protect\citeauthoryear{Ahmed, Hedar, and Ibrahim}{Ahmed
  et~al\mbox{.}}{2014}]%
        {ahmed2014predicting}
\bibfield{author}{\bibinfo{person}{Mostafa~M Ahmed}, \bibinfo{person}{Abdel
  Rahman~M Hedar}, {and} \bibinfo{person}{Hosny~M Ibrahim}.}
  \bibinfo{year}{2014}\natexlab{}.
\newblock \showarticletitle{Predicting bug category based on analysis of
  software repositories}. In \bibinfo{booktitle}{\emph{Proceedings of the 2nd
  International Conference on Research in Science, Engineering and Technology
  (ICRSET’2014)}}. \bibinfo{pages}{44--53}.
\newblock


\bibitem[\protect\citeauthoryear{Alenezi, Magel, and Banitaan}{Alenezi
  et~al\mbox{.}}{2013}]%
        {alenezi2013efficient}
\bibfield{author}{\bibinfo{person}{Mamdouh Alenezi}, \bibinfo{person}{Kenneth
  Magel}, {and} \bibinfo{person}{Shadi Banitaan}.}
  \bibinfo{year}{2013}\natexlab{}.
\newblock \showarticletitle{Efficient Bug Triaging Using Text Mining.}
\newblock \bibinfo{journal}{\emph{J. Softw.}} \bibinfo{volume}{8},
  \bibinfo{number}{9} (\bibinfo{year}{2013}), \bibinfo{pages}{2185--2190}.
\newblock


\bibitem[\protect\citeauthoryear{Alkhazi, DiStasi, Aljedaani, Alrubaye, Ye, and
  Mkaouer}{Alkhazi et~al\mbox{.}}{2020}]%
        {alkhazi2020learning}
\bibfield{author}{\bibinfo{person}{Bader Alkhazi}, \bibinfo{person}{Andrew
  DiStasi}, \bibinfo{person}{Wajdi Aljedaani}, \bibinfo{person}{Hussein
  Alrubaye}, \bibinfo{person}{Xin Ye}, {and} \bibinfo{person}{Mohamed~Wiem
  Mkaouer}.} \bibinfo{year}{2020}\natexlab{}.
\newblock \showarticletitle{Learning to rank developers for bug report
  assignment}.
\newblock \bibinfo{journal}{\emph{Applied Soft Computing}}
  \bibinfo{volume}{95} (\bibinfo{year}{2020}), \bibinfo{pages}{106667}.
\newblock


\bibitem[\protect\citeauthoryear{Antoniol, Ayari, Di~Penta, Khomh, and
  Gu\'{e}h\'{e}neuc}{Antoniol et~al\mbox{.}}{2008}]%
        {antonio}
\bibfield{author}{\bibinfo{person}{Giuliano Antoniol}, \bibinfo{person}{Kamel
  Ayari}, \bibinfo{person}{Massimiliano Di~Penta}, \bibinfo{person}{Foutse
  Khomh}, {and} \bibinfo{person}{Yann-Ga\"{e}l Gu\'{e}h\'{e}neuc}.}
  \bibinfo{year}{2008}\natexlab{}.
\newblock \showarticletitle{Is It a Bug or an Enhancement? A Text-Based
  Approach to Classify Change Requests}. In
  \bibinfo{booktitle}{\emph{Proceedings of the 2008 Conference of the Center
  for Advanced Studies on Collaborative Research: Meeting of Minds}} (Ontario,
  Canada) \emph{(\bibinfo{series}{CASCON '08})}.
  \bibinfo{publisher}{Association for Computing Machinery},
  \bibinfo{address}{New York, NY, USA}, Article \bibinfo{articleno}{23},
  \bibinfo{numpages}{15}~pages.
\newblock
\showISBNx{9781450378826}
\urldef\tempurl%
\url{https://doi.org/10.1145/1463788.1463819}
\showDOI{\tempurl}


\bibitem[\protect\citeauthoryear{Anvik, Hiew, and Murphy}{Anvik
  et~al\mbox{.}}{2006}]%
        {anvik2006should}
\bibfield{author}{\bibinfo{person}{John Anvik}, \bibinfo{person}{Lyndon Hiew},
  {and} \bibinfo{person}{Gail~C Murphy}.} \bibinfo{year}{2006}\natexlab{}.
\newblock \showarticletitle{Who should fix this bug?}. In
  \bibinfo{booktitle}{\emph{Proceedings of the 28th international conference on
  Software engineering}}. \bibinfo{pages}{361--370}.
\newblock


\bibitem[\protect\citeauthoryear{Borges, Hora, and Valente}{Borges
  et~al\mbox{.}}{2016}]%
        {repopopularity}
\bibfield{author}{\bibinfo{person}{Hudson Borges}, \bibinfo{person}{Andre
  Hora}, {and} \bibinfo{person}{Marco~Tulio Valente}.}
  \bibinfo{year}{2016}\natexlab{}.
\newblock \showarticletitle{Understanding the Factors That Impact the
  Popularity of GitHub Repositories}. In \bibinfo{booktitle}{\emph{2016 IEEE
  International Conference on Software Maintenance and Evolution (ICSME)}}.
  \bibinfo{pages}{334--344}.
\newblock
\urldef\tempurl%
\url{https://doi.org/10.1109/ICSME.2016.31}
\showDOI{\tempurl}


\bibitem[\protect\citeauthoryear{Cass}{Cass}{2020}]%
        {languageranking}
\bibfield{author}{\bibinfo{person}{Stephen Cass}.}
  \bibinfo{year}{2020}\natexlab{}.
\newblock \showarticletitle{The top programming languages: Our latest rankings
  put Python on top-again - [Careers]}.
\newblock \bibinfo{journal}{\emph{IEEE Spectrum}} \bibinfo{volume}{57},
  \bibinfo{number}{8} (\bibinfo{year}{2020}), \bibinfo{pages}{22--22}.
\newblock
\urldef\tempurl%
\url{https://doi.org/10.1109/MSPEC.2020.9150550}
\showDOI{\tempurl}


\bibitem[\protect\citeauthoryear{Ded{\'\i}k and Rossi}{Ded{\'\i}k and
  Rossi}{2016}]%
        {dedik2016automated}
\bibfield{author}{\bibinfo{person}{V{\'a}clav Ded{\'\i}k} {and}
  \bibinfo{person}{Bruno Rossi}.} \bibinfo{year}{2016}\natexlab{}.
\newblock \showarticletitle{Automated bug triaging in an industrial context}.
  In \bibinfo{booktitle}{\emph{2016 42th Euromicro Conference on Software
  Engineering and Advanced Applications (SEAA)}}. IEEE,
  \bibinfo{pages}{363--367}.
\newblock


\bibitem[\protect\citeauthoryear{Devlin, Chang, Lee, and Toutanova}{Devlin
  et~al\mbox{.}}{2018a}]%
        {bert}
\bibfield{author}{\bibinfo{person}{Jacob Devlin}, \bibinfo{person}{Ming{-}Wei
  Chang}, \bibinfo{person}{Kenton Lee}, {and} \bibinfo{person}{Kristina
  Toutanova}.} \bibinfo{year}{2018}\natexlab{a}.
\newblock \showarticletitle{{BERT:} Pre-training of Deep Bidirectional
  Transformers for Language Understanding}.
\newblock \bibinfo{journal}{\emph{CoRR}}  \bibinfo{volume}{abs/1810.04805}
  (\bibinfo{year}{2018}).
\newblock
\showeprint[arxiv]{1810.04805}
\urldef\tempurl%
\url{http://arxiv.org/abs/1810.04805}
\showURL{%
\tempurl}


\bibitem[\protect\citeauthoryear{Devlin, Chang, Lee, and Toutanova}{Devlin
  et~al\mbox{.}}{2018b}]%
        {devlin2018bert}
\bibfield{author}{\bibinfo{person}{Jacob Devlin}, \bibinfo{person}{Ming-Wei
  Chang}, \bibinfo{person}{Kenton Lee}, {and} \bibinfo{person}{Kristina
  Toutanova}.} \bibinfo{year}{2018}\natexlab{b}.
\newblock \showarticletitle{Bert: Pre-training of deep bidirectional
  transformers for language understanding}.
\newblock \bibinfo{journal}{\emph{arXiv preprint arXiv:1810.04805}}
  (\bibinfo{year}{2018}).
\newblock


\bibitem[\protect\citeauthoryear{Fan, Yu, Yin, Wang, and Wang}{Fan
  et~al\mbox{.}}{2017}]%
        {fan}
\bibfield{author}{\bibinfo{person}{Qiang Fan}, \bibinfo{person}{Yue Yu},
  \bibinfo{person}{Gang Yin}, \bibinfo{person}{Tao Wang}, {and}
  \bibinfo{person}{Huaimin Wang}.} \bibinfo{year}{2017}\natexlab{}.
\newblock \showarticletitle{Where Is the Road for Issue Reports Classification
  Based on Text Mining?}. In \bibinfo{booktitle}{\emph{2017 ACM/IEEE
  International Symposium on Empirical Software Engineering and Measurement
  (ESEM)}}. \bibinfo{pages}{121--130}.
\newblock
\urldef\tempurl%
\url{https://doi.org/10.1109/ESEM.2017.19}
\showDOI{\tempurl}


\bibitem[\protect\citeauthoryear{Fazayeli, Syed-Mohamad, and {Md
  Akhir}}{Fazayeli et~al\mbox{.}}{2019}]%
        {FAZAYELI2019585}
\bibfield{author}{\bibinfo{person}{Hassan Fazayeli},
  \bibinfo{person}{Sharifah~Mashita Syed-Mohamad}, {and}
  \bibinfo{person}{Nur~Shazwani {Md Akhir}}.} \bibinfo{year}{2019}\natexlab{}.
\newblock \showarticletitle{Towards Auto-labelling Issue Reports for Pull-Based
  Software Development using Text Mining Approach}.
\newblock \bibinfo{journal}{\emph{Procedia Computer Science}}
  \bibinfo{volume}{161} (\bibinfo{year}{2019}), \bibinfo{pages}{585--592}.
\newblock
\showISSN{1877-0509}
\urldef\tempurl%
\url{https://doi.org/10.1016/j.procs.2019.11.160}
\showDOI{\tempurl}
\newblock
\shownote{The Fifth Information Systems International Conference, 23-24 July
  2019, Surabaya, Indonesia.}


\bibitem[\protect\citeauthoryear{Hu}{Hu}{2020}]%
        {attention_nlp}
\bibfield{author}{\bibinfo{person}{Dichao Hu}.}
  \bibinfo{year}{2020}\natexlab{}.
\newblock \showarticletitle{An Introductory Survey on Attention Mechanisms in
  NLP Problems}. In \bibinfo{booktitle}{\emph{Intelligent Systems and
  Applications}}, \bibfield{editor}{\bibinfo{person}{Yaxin Bi},
  \bibinfo{person}{Rahul Bhatia}, {and} \bibinfo{person}{Supriya Kapoor}}
  (Eds.). \bibinfo{publisher}{Springer International Publishing},
  \bibinfo{address}{Cham}, \bibinfo{pages}{432--448}.
\newblock
\showISBNx{978-3-030-29513-4}


\bibitem[\protect\citeauthoryear{Joulin, Grave, Bojanowski, and Mikolov}{Joulin
  et~al\mbox{.}}{2016}]%
        {joulin2016bag}
\bibfield{author}{\bibinfo{person}{Armand Joulin}, \bibinfo{person}{Edouard
  Grave}, \bibinfo{person}{Piotr Bojanowski}, {and} \bibinfo{person}{Tomas
  Mikolov}.} \bibinfo{year}{2016}\natexlab{}.
\newblock \showarticletitle{Bag of tricks for efficient text classification}.
\newblock \bibinfo{journal}{\emph{arXiv preprint arXiv:1607.01759}}
  (\bibinfo{year}{2016}).
\newblock


\bibitem[\protect\citeauthoryear{Kalliamvakou, Gousios, Blincoe, Singer,
  German, and Damian}{Kalliamvakou et~al\mbox{.}}{2014}]%
        {kalliamvakou2014promises}
\bibfield{author}{\bibinfo{person}{Eirini Kalliamvakou},
  \bibinfo{person}{Georgios Gousios}, \bibinfo{person}{Kelly Blincoe},
  \bibinfo{person}{Leif Singer}, \bibinfo{person}{Daniel~M German}, {and}
  \bibinfo{person}{Daniela Damian}.} \bibinfo{year}{2014}\natexlab{}.
\newblock \showarticletitle{The promises and perils of mining github}. In
  \bibinfo{booktitle}{\emph{Proceedings of the 11th working conference on
  mining software repositories}}. \bibinfo{pages}{92--101}.
\newblock


\bibitem[\protect\citeauthoryear{Kallis, {Di Sorbo}, Canfora, and
  Panichella}{Kallis et~al\mbox{.}}{2021}]%
        {KALLIS2021102598}
\bibfield{author}{\bibinfo{person}{Rafael Kallis}, \bibinfo{person}{Andrea {Di
  Sorbo}}, \bibinfo{person}{Gerardo Canfora}, {and} \bibinfo{person}{Sebastiano
  Panichella}.} \bibinfo{year}{2021}\natexlab{}.
\newblock \showarticletitle{Predicting issue types on GitHub}.
\newblock \bibinfo{journal}{\emph{Science of Computer Programming}}
  \bibinfo{volume}{205} (\bibinfo{year}{2021}), \bibinfo{pages}{102598}.
\newblock
\showISSN{0167-6423}
\urldef\tempurl%
\url{https://doi.org/10.1016/j.scico.2020.102598}
\showDOI{\tempurl}


\bibitem[\protect\citeauthoryear{Kevic, M{\"u}ller, Fritz, and Gall}{Kevic
  et~al\mbox{.}}{2013}]%
        {kevic2013collaborative}
\bibfield{author}{\bibinfo{person}{Katja Kevic}, \bibinfo{person}{Sebastian~C
  M{\"u}ller}, \bibinfo{person}{Thomas Fritz}, {and} \bibinfo{person}{Harald~C
  Gall}.} \bibinfo{year}{2013}\natexlab{}.
\newblock \showarticletitle{Collaborative bug triaging using textual
  similarities and change set analysis}. In \bibinfo{booktitle}{\emph{2013 6th
  International Workshop on Cooperative and Human Aspects of Software
  Engineering (CHASE)}}. IEEE, \bibinfo{pages}{17--24}.
\newblock


\bibitem[\protect\citeauthoryear{Lin, Shu, Yang, Hu, and Wang}{Lin
  et~al\mbox{.}}{2009}]%
        {bug_assignment_chinese}
\bibfield{author}{\bibinfo{person}{Zhongpeng Lin}, \bibinfo{person}{Fengdi
  Shu}, \bibinfo{person}{Ye Yang}, \bibinfo{person}{Chenyong Hu}, {and}
  \bibinfo{person}{Qing Wang}.} \bibinfo{year}{2009}\natexlab{}.
\newblock \showarticletitle{An empirical study on bug assignment automation
  using Chinese bug data}. In \bibinfo{booktitle}{\emph{2009 3rd International
  Symposium on Empirical Software Engineering and Measurement}}. IEEE,
  \bibinfo{pages}{451--455}.
\newblock


\bibitem[\protect\citeauthoryear{Liu, Ott, Goyal, Du, Joshi, Chen, Levy, Lewis,
  Zettlemoyer, and Stoyanov}{Liu et~al\mbox{.}}{2019}]%
        {roberta}
\bibfield{author}{\bibinfo{person}{Yinhan Liu}, \bibinfo{person}{Myle Ott},
  \bibinfo{person}{Naman Goyal}, \bibinfo{person}{Jingfei Du},
  \bibinfo{person}{Mandar Joshi}, \bibinfo{person}{Danqi Chen},
  \bibinfo{person}{Omer Levy}, \bibinfo{person}{Mike Lewis},
  \bibinfo{person}{Luke Zettlemoyer}, {and} \bibinfo{person}{Veselin
  Stoyanov}.} \bibinfo{year}{2019}\natexlab{}.
\newblock \bibinfo{title}{RoBERTa: A Robustly Optimized BERT Pretraining
  Approach}.
\newblock
\newblock
\showeprint[arxiv]{1907.11692}~[cs.CL]


\bibitem[\protect\citeauthoryear{Matter, Kuhn, and Nierstrasz}{Matter
  et~al\mbox{.}}{2009}]%
        {matter2009assigning}
\bibfield{author}{\bibinfo{person}{Dominique Matter}, \bibinfo{person}{Adrian
  Kuhn}, {and} \bibinfo{person}{Oscar Nierstrasz}.}
  \bibinfo{year}{2009}\natexlab{}.
\newblock \showarticletitle{Assigning bug reports using a vocabulary-based
  expertise model of developers}. In \bibinfo{booktitle}{\emph{2009 6th IEEE
  international working conference on mining software repositories}}. IEEE,
  \bibinfo{pages}{131--140}.
\newblock


\bibitem[\protect\citeauthoryear{Murphy and Cubranic}{Murphy and
  Cubranic}{2004}]%
        {murphy2004automatic}
\bibfield{author}{\bibinfo{person}{G Murphy} {and} \bibinfo{person}{Davor
  Cubranic}.} \bibinfo{year}{2004}\natexlab{}.
\newblock \showarticletitle{Automatic bug triage using text categorization}. In
  \bibinfo{booktitle}{\emph{Proceedings of the Sixteenth International
  Conference on Software Engineering \& Knowledge Engineering}}. Citeseer,
  \bibinfo{pages}{1--6}.
\newblock


\bibitem[\protect\citeauthoryear{Sajedi~Badashian, Hindle, and
  Stroulia}{Sajedi~Badashian et~al\mbox{.}}{2015}]%
        {crowdsourcedtriage}
\bibfield{author}{\bibinfo{person}{Ali Sajedi~Badashian},
  \bibinfo{person}{Abram Hindle}, {and} \bibinfo{person}{Eleni Stroulia}.}
  \bibinfo{year}{2015}\natexlab{}.
\newblock \showarticletitle{Crowdsourced bug triaging}. In
  \bibinfo{booktitle}{\emph{2015 IEEE International Conference on Software
  Maintenance and Evolution (ICSME)}}. \bibinfo{pages}{506--510}.
\newblock
\urldef\tempurl%
\url{https://doi.org/10.1109/ICSM.2015.7332503}
\showDOI{\tempurl}


\bibitem[\protect\citeauthoryear{Vaswani, Shazeer, Parmar, Uszkoreit, Jones,
  Gomez, Kaiser, and Polosukhin}{Vaswani et~al\mbox{.}}{2017}]%
        {vaswani2017attention}
\bibfield{author}{\bibinfo{person}{Ashish Vaswani}, \bibinfo{person}{Noam
  Shazeer}, \bibinfo{person}{Niki Parmar}, \bibinfo{person}{Jakob Uszkoreit},
  \bibinfo{person}{Llion Jones}, \bibinfo{person}{Aidan~N. Gomez},
  \bibinfo{person}{Lukasz Kaiser}, {and} \bibinfo{person}{Illia Polosukhin}.}
  \bibinfo{year}{2017}\natexlab{}.
\newblock \bibinfo{title}{Attention Is All You Need}.
\newblock
\newblock
\showeprint[arxiv]{1706.03762}~[cs.CL]


\end{thebibliography}
\end{document}